\documentclass[prd,nofootinbib,aps,twocolumn,floats,floatfix,
superscriptaddress,amsmath,amssymb,secnumarabic]{revtex4}
\usepackage{natbib}
\usepackage{moreverb}
\usepackage{graphicx}
\usepackage{amsmath}
\usepackage{amssymb}
\usepackage{float}
\usepackage{slashed}
\usepackage{color}
\usepackage[utf8]{inputenc}
\usepackage{hyperref}

\newcommand{\be}{\begin{equation}}
\newcommand{\ee}{\end{equation}}
\newcommand{\ba}{\begin{array}}
\newcommand{\ea}{\end{array}}
\newcommand{\bea}{\begin{eqnarray}}
\newcommand{\eea}{\end{eqnarray}}
\newcommand{\sss}{\scriptscriptstyle}
\def\sfrac#1#2{{\textstyle{#1\over #2}}}

\begin{document}

\title{Minimal nonabelian model of atomic dark matter}
\author{Jeremie Choquette\footnote{jeremie.choquette@physics.mcgill.ca}}
\affiliation{Department of Physics, McGill University,
3600 Rue University, Montr\'eal, Qu\'ebec, Canada H3A 2T8}
\author{James M.\ Cline\footnote{jcline@physics.mcgill.ca}}
\affiliation{Department of Physics, McGill University,
3600 Rue University, Montr\'eal, Qu\'ebec, Canada H3A 2T8}
\affiliation{Niels Bohr International Academy and Discovery Center,
Niels Bohr Institute, University of Copenhagen,
Blegdamsvej 17, DK-2100 Copenhagen \O, Denmark}
\begin{abstract}
A dark sector resembling the standard model, where the abundance
of matter is explained by baryon and lepton asymmetries, and stable
constituents bind to form atoms, is a theoretically appealing 
possibility.  We show that a minimal model with a hidden SU(2)
gauge symmetry broken
to U(1), with a Dirac fermion doublet, suffices to realize 
this scenario.  Supplemented with a dark Higgs doublet that gets no
VEV, we readily achieve the dark matter asymmetry through
leptogenesis.  The model can simultaneously have three portals to
the standard model, through the Higgs, nonabelian kinetic mixing, 
and the heavy neutrino, with interesting phenomenology for
direct and collider searches, as well as cosmologically
relevant DM self-interactions.  Exotic bound states consisting of
two fermions and a doubly-charged vector boson can exist in one
phase of the theory.
\end{abstract}
\maketitle

Dark matter (DM) from a hidden sector has been a popular alternative to
supersymmetric weakly interacting massive particles in recent years
\cite{Pospelov:2007mp,ArkaniHamed:2008qn}.  A widely studied example
is dark atoms, where the DM consists of two species with opposite
charges under an unbroken U(1)$_h$ hidden sector gauge symmetry
\cite{Feng:2009mn,Kaplan:2009de,Kaplan:2011yj,Fan:2013tia,Fan:2013yva,Foot:2014mia,Garcia:2015toa}.  
This class of models
presents rich possibilities for direct detection
\cite{Fan:2013bea,Foot:2014osa,Frandsen:2014lfa,Pearce:2015zca},
as well as cosmological imprints
\cite{CyrRacine:2012fz,Cyr-Racine:2013fsa,Cline:2013pca,
Cline:2013zca,Foot:2014uba,Petraki:2014uza}.
If the hidden photon has kinetic mixing with the normal
photon, the dark constituents acquire electric millicharges
\cite{Holdom:1985ag}, leading
to further constraints and prospects for detection 
\cite{Cline:2012is,Cline:2012ei,Cline:2014eaa}.  

Simplified models of atomic dark matter are easy to construct,
consisting of just two fermions and the gauge boson in the hidden
sector, but such examples are necessarily incomplete descriptions
of the new physics required.  First, it is desirable for the DM to
be asymmetric, otherwise the long-range U(1)$_h$ interaction would
leave too small a relic abundance unless the DM mass exceeds 
$\sim 400$ GeV \cite{Ackerman:mha}\footnote{
For lower masses the DM
self-interactions violate bounds from structure formation.
This argument assumes that the DM remains ionized, which
turns out to be valid for the gauge coupling strength needed
to get the right relic density from thermal freezeout.}.  
Simplified models do not explain the origin of the asymmetry.
Second, the U(1)$_h$ gauge interaction leads to a Landau pole at
high energies, so it would be desirable to find a more UV-complete
version of the theory.  Third, dark constituent millicharges
greater than $\sim 10^{-7}e$ (of interest for collider searches)
require the atomic constituents to be 
nearly equal in mass, which is a rather ad hoc requirement in the
simplified models.    
In this work we present a model that is still
relatively simple, but addresses both of these issues, and makes a number
of interesting experimental predictions.  It relies upon breaking a
nonabelian (hence asymptotically free) gauge symmetry SU(2)$_h$ down
to U(1)$_h$ to explain the origin of the massless dark photon.  The
approximate equality of the dark consituents, if desired,
can be explained as a remnant of the gauge symmetry.

There have been many proposals for mechanisms that link the
asymmetries of the hidden and visible sectors.  In general, they tend
to be complicated.  A notable exception is to use the
out-of-equilibrium decays of heavy neutrinos to generate both asymmetries via
leptogenesis and its analog in the hidden sector 
\cite{An:2009vq,Chun:2010hz,Chun:2011cc,Arina:2011cu,Kaplan:2011yj,
Falkowski:2011xh,Feng:2013zda}. We adopt this approach here.

The model presents opportunities for direct detection, either through
Higgs portal interactions or nonabelian gauge kinetic mixing. The
latter can arise through a dimension-5 operator involving the triplet
Higgs field that breaks the SU(2)$_h$ gauge symmetry
\cite{Chen:2009ab}.  This results in electric millicharges for the
dark matter constituents, that normally must be very small to avoid 
direct detection, but can be sizable if the dark constituents have
equal mass, which is a symmetry limit of the theory presented here.  Moreover the
self-interactions of the dark atoms can be of the right magnitude for
addressing  problems of small-scale structure formation in standard
noninteracting  $\Lambda$CDM cosmology.

In the following we introduce the model (section \ref{model}) and
then estimate the dark matter and baryon asymmetries that can arise in
a generic scenario for leptogenesis (section \ref{lepto}).  Limits from direct searches
are worked out in section \ref{ddsect}. In section \ref{h2sect} we consider the region of parameter space in which the vector bosons are stable, leading to a markedly different dark sector. In sect.\ \ref{other}
we discuss constraints pertaining to the ionization fraction, 
dark atom self-interactions, and searches for millicharged
particles.   
Conclusions are given in sect.\ \ref{conclusion}.

\section{The model}
\label{model} 

 The new-physics content of the model (summarized in 
table~\ref{tab:model}) is a
hidden SU(2)$_h$
gauge boson $B_\mu$ with field strength $B^{a}_{\mu\nu}$, a 
real scalar triplet $\phi$ that spontaneously
breaks SU(2)$_h$ by getting a VEV, a scalar doublet $\eta$ that
does not get a VEV, 
two Weyl fermion doublets $\psi_i^\alpha$ (with gauge index $\alpha$ and
flavor index $i$)
 and the heavy right-handed neutrinos $N_j$ that also
interact in the usual way with the standard model neutrinos.
An even number of fermion doublets is required to 
avoid Witten's global SU(2)
anomaly \cite{Witten:1982fp}.  They can be combined into a
Dirac doublet fermion $\Psi = (\psi_{1L}, \psi^{c}_{2R})$ where the
conjugate is defined as $\psi^{c}_{2R} = \sigma_2\tau_2
(\psi_{2L})^*$, {\it i.e.,} the epsilon tensor is applied both to the
spin and to the SU(2)$_h$ gauge indices.   
Without loss of generality the VEV of $\phi$ can be
rotated to the 3rd component,  $\langle
\phi^a\rangle=(0,0,\sigma)$.

\begin{table}[H]
\centerline{
\begin{tabular}{|l|c|c|c|c|c|c|}
\hline
${\hbox{particle}_{\phantom{|}}^{\phantom{|}}\atop \hbox{VEV}^{\phantom{|}}}$&$B^{0,++,--}$&\parbox[c]{1.8cm}{\vspace{3pt}
\centering $\phi^a\rightarrow$\\ $(0,0,\sigma+\phi)$}&
$\eta^{+,-}$&$\Psi_1^{+,-}$&$\Psi_2^{+,-}$ & $N_j$\\
\hline
Spin&1&0&0&$\frac12$&$\frac12$&$\frac12$\\
SU(2)$_h$  & 3 & 3 & 2 & 2 & 2 & 1\\
U$(1)_h$ &$0,\,+2,\,-2$&0&$+1,\,-1$&$+1,\,-1$&$+1,\,-1$&0\\
\hline
\end{tabular}
}
\caption{New particle content in the model, showing the Lorentz,
hidden SU(2) and hidden U(1) (after breaking of SU(2)$_h\to$U(1)$_h$)
quantum numbers.}
\label{tab:model}
\end{table}

The relevant terms in the Lagrangian are
\bea
\mathcal{L}= &-&\frac{1}{4}B^a_{\mu\nu}B_a^{\mu\nu}+\frac12(D_\mu \phi)^2
-{1\over\Lambda} \phi^a\,B^a_{\mu\nu}Y^{\mu\nu}\\
&+& \bar\Psi(i\slashed{D}-m_\psi)\Psi - \bar\Psi \,(y_1 +
iy_2\gamma_5)
	(\vec\phi\cdot\vec\tau)\, \Psi 
\\
 &-& |D_\mu\eta|^2 - V(H,\phi,\eta)\nonumber\\ &-& 
(\bar\psi^i_{L} \eta)\,y_{\psi}^{ij} P_R N_j  
+{\rm h.c.}\nonumber
\label{Eq:Lagrangian}
\eea
where the covariant derivative is 
$D_\mu\phi^a =\partial_\mu\phi^a - g\,\epsilon_{abc}\,B^b_\mu\phi^c$
or $D_\mu\Psi = (\partial_\mu -
i(g/2)\vec B_\mu\cdot\vec\tau)\Psi$, 
$g$ is the SU(2)$_h$ gauge coupling, and $Y_{\mu\nu}$
is the Standard Model hypercharge gauge field strength.
In the Yukawa interactions with the sterile neutrino we use the Weyl fermion notation since the
analogy to leptogenesis via neutrino physics is more clear in 
this way. 

The triplet scalar VEV breaks SU(2)$_h$ to U(1)$_h$, mediated by the 
massless gauge boson $B_3^\mu$, while $B^{\pm\!\pm} = (B^1\pm
iB^2)/\sqrt{2}$ obtain mass $m_B = g\sigma$.  The upper and lower
components $\Psi_{1,2}$ of the fermion doublet are also
charged under the U(1)$_h$ (with half the charge of $B^{\pm\!\pm}$).
Their masses are split by the Yukawa interaction, $m_{1,2} = ((m_\psi
\pm y_1\sigma)^2 + (y_2\sigma)^2)^{1/2}$.   We used the freedom to
perform a chiral rotation on $\Psi$ so that $m_\psi$ is real (has no
$\gamma_5$ component).  

In section~\ref{decays} we discuss the decay of the scalars 
through $\eta\rightarrow\psi\nu$ via 
the dimension-5 operator 
\be
	\bar\psi_{i,L} \eta\,\,y_\psi^{ij}M_j^{-1}y_{\nu}^{jk} P_R\,(H^T \bar
	L_k^T) + {\rm h.c.}
\label{dim5op}
\ee
where $M_j$ is the heavy neutrino mass (in a basis where its mass
matrix is diagonal), $y_\nu$ is the neutrino 
Yukawa matrix, $H$ is the SM Higgs doublet, and $L_k$ are the lepton
doublets.
 
 We will initially consider the case where decays  
$\Psi_1\to B^{++} \Psi_2$ are not
kinematically allowed.  They would lead to a dark sector consisting
of stable $B^{++}$ vector bosons and $\Psi_2^-$ fermions. (The
alternative case in which these decays are allowed is considered in 
section \ref{h2sect}.) This leaves  two species of stable dark matter, 
the Dirac fermions $\Psi_{1} = (\psi^1_{1L},\psi^{2c}_{2R})^T$ and 
$\Psi_{2} = (\psi^2_{1L},\psi^{1c}_{2R})^T$
with charges $\pm 1$ under the unbroken U(1)$_h$. 
The long-range force mediated by the dark photon $B_3\equiv \gamma'$ 
causes the symmetric component of the DM densities to be at least
partially depleted by annihilations, and the asymmetric components
of $\Psi_{1,2}$ to bind into dark atoms.  The efficiency of these
processes depends upon the gauge coupling $g$ and the dark atom
mass $m_{\bf H}$, as we will discuss in section \ref{other}.  
 
For simplicity we impose a softly broken  U(1) symmetry under which
$\psi_i \to e^{i\theta}\psi_i$, $\eta \to e^{i\theta}\eta$, which
forbids the interactions  $(\bar\psi_i\tilde\eta)N_j$, with
$\tilde\eta = \tau_2\eta^*$. The symmetry is broken by the Dirac mass
term, which takes the form
\be
	-m_\psi(\bar\psi_{2R}^{2c}\,\psi^1_{1L} + 
		\bar\psi_{2R}^{1c}\,\psi^2_{1L}) + 
	{\rm h.c.}
\ee
If the symmetry were exact, then the subsequent decays $\eta\to\psi$
mediated by $N_i$ would completely erase any produced DM asymmetry. 
However the chirality flips induced  by the mass term prevent this
erasure, as we will explain in more detail in section \ref{lepto}. 
There is an unbroken discrete $Z_2$ remnant of this symmetry,
where $\psi_i\to -\psi_i$ and $\eta\to -\eta$, that ensures the
stability of the dark matter.

\begin{figure}[t]
\centerline{
\includegraphics[width=0.7\columnwidth]{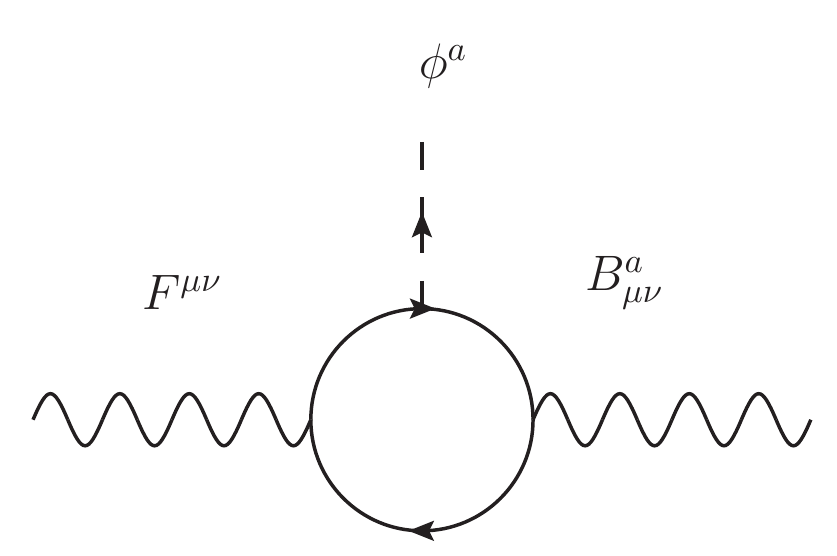}
}
\caption{Loop contribution to the nonabelian kinetic mixing operator.}
\label{fig:loop}
\end{figure}

The potential $V$
is assumed to give rise to the VEV of $\phi$
and it generically also includes the
Higgs portal coupling $\frac12\lambda_{h\phi}|H|^2 \phi^2$.
Once $\phi$ gets its VEV, the nonabelian kinetic mixing operator
can
be written as
\be
	-\frac12\sin\tilde\epsilon\, B^3_{\mu\nu}Y^{\mu\nu}
\label{ngkm}
\ee
where $\sin\tilde\epsilon = 2\sigma/\Lambda$.
It could arise from integrating out a heavy vector-like fermion
$\chi$ that carries hypercharge and transforms as a doublet
under SU(2)$_h$.  The interaction $y_\chi\bar\chi \phi_a\sigma_a \chi$
leads to the diagram in fig.\ \ref{fig:loop}, implying
$\Lambda^{-1} \sim g g_1 y_\chi /m_{\chi}$ where $g_1$ is the
hypercharge coupling.  
The kinetic mixing gives rise to electric millicharges 
$\pm\tilde\epsilon g\equiv \pm \epsilon e$ 
for the fermions $\Psi_{1,2}$.  This or alternatively the
Higgs portal interaction allows for
direct detection of the dark atoms, as we discuss in section 
\ref{ddsect}.

\section{Origin of dark matter asymmetry}
\label{lepto}

Our setup allows for heavy neutrinos to decay in a CP-violating 
manner into an excess of dark matter versus its antiparticles
in close analogy to leptogenesis.  The structure of Yukawa couplings
is similar to that of neutrinos except that we have only two light 
fermionic DM
species, $\Psi_{1,2}$ as compared to the three light neutrinos.
The dark Higgs doublet $\eta$ does not have a VEV, so it also
gets an asymmetry, which will be determined by those in 
$\Psi_i$.

The asymmetry in the decay of the $j$th heavy neutrino into 
$\psi_i^*\eta$ versus $\psi_i \eta^*$ (recall that
$\psi_i$ denotes the Weyl doublet states) is given by
\bea
\label{epsij}
	\epsilon_{\psi}^{ji} &=& {\Gamma(N_j\to \psi_i^*\eta) - 
	\Gamma(N_j\to \psi_i\eta^*)\over \Gamma(N_i\to {\rm any})}
	\\
	&=& {1\over 8\pi}
	\sum_{k\neq j}\Bigg[
{{\rm Im}\left[ (y_\psi^\dagger y_\psi)_{kj}y_{\psi}^{ik}y_\psi^{ij*}
	\right]\over
(y_\psi^\dagger y_\psi 	+  y_\nu^\dagger y_\nu)_{jj}}\,
 g(M_k^2/M_j^2)\nonumber\\
&\quad& 
\quad\quad\ +\ \ {{\rm Im}\left[ (y_\nu^\dagger y_\nu)_{kj}
y_\psi^{ik}y_\psi^{ij*}
	\right]\over
(y_\psi^\dagger y_\psi +  y_\nu^\dagger y_\nu)_{jj}}\,
 g'(M_k^2/M_j^2)\Bigg]\nonumber
\eea 
where $g(x)=\sqrt{x}\left[1/(1-x)+1-(1+x)\ln (1+1/x)\right]$
and $g'(x)=\sqrt{x}/(1-x)$.  
This differs from the standard leptogenesis expression because 
the denominator must take into account decays of $N_j$ both into
neutrinos and dark matter, and there is a mixed term of order 
$y_\nu^2\, y_\psi^2$ from the self-energy correction of $N_j$ by the
SM Yukawa interaction.

For definiteness, we will focus on decay of the lightest heavy
neutrino $N_1$.  In the simplest scenario of leptogenesis, where
$M_1\ll M_{2,3}$ and the reheat temperature is in between,  $M_1 <
T_{rh} < M_{2,3}$, this is the only  relevant decay since the heavier
neutrinos are not present.  In this case the functions in 
eq.\ (\ref{epsij}) can be approximated as 
$g\cong -3/2\sqrt{x}$ and $g'\cong -1/\sqrt{x}$ with
$x = (M_2/M_1)^2\gg 1$.

Initially, we can expect independent asymmetries $Y_{1,2}$ for
$\psi_1$ and $\psi_2$, where $Y_i = (n_{\psi_i} - n_{\bar\psi_i})/s$
is the dark matter to entropy ratio, since $\epsilon_\psi^{11}
 \neq
\epsilon_\psi^{12}$. 
However the Dirac mass term takes the
form $\psi_{1}^T\sigma_2\tau_2\psi_2$, which implies that mass effects
will cause the asymmetries of $\psi_1$ and $\psi_2$ to become equal
and opposite.  This projects the net asymmetry of the fermions onto
the difference between the initial ones, $Y_\psi = Y_1 - Y_2$, at
temperatures where the helicity-flipping interactions due to $m_\psi$
come into equilibrium.  

On the other hand, the $\eta$ boson  gets a different asymmetry,
proportional to  $\epsilon_\psi^{11} +
\epsilon_\psi^{12} $.  Eventually it will
decay into $\psi_i$. For simplicity, we consider
the case $\epsilon_\psi^{11}\sim-\epsilon_\psi^{12}$. 
Then not only does the initial asymmetry in
$\eta$ tend to be small, but so also is its contribution to the
final asymmetry in $\psi_i$, and 
we can estimate the net asymmetry in $\psi$ from $N_1$
decays as 
\bea
	\epsilon_{\psi 1} &\sim& \epsilon_\psi^{11} 
	-\epsilon_\psi^{12} \sim 2\epsilon_\psi^{11}
\label{epsj}
\eea
The sign difference is
in contrast to the CP asymmetries for decays into neutrinos,
$\epsilon_{\nu 1} = \sum_i \epsilon_{\nu}^{1i}$ familiar from 
leptogenesis.

\subsection{Dark matter asymmetry estimate}

The initial asymmetries depend upon an efficiency factor $\kappa_\psi$
that quantifies the amount of washout (see for example 
\cite{Buchmuller:2004nz} for a review).  The contribution from $N_1$ 
decay is 
\be	Y_\psi = {45\over \pi^4}{\epsilon_{\psi 1} \kappa_\psi\over g_*}
\label{Yieq}
\ee
where $\kappa_\psi\cong {\rm min}(0.25\,(m_*/\tilde m_{\psi 1})^{1.1},\, 1)$ with
$\tilde m_{\psi 1} = 2(y_\psi^\dagger y_\psi)_{11} v^2/M_1$, $m_*=10^{-3}{\rm\, eV}$ and 
$v=174$ GeV.
The Higgs
VEV $v$ has no direct physical relevance for the dark matter abundance, but
$\tilde m_{\psi 1}/m_*$ gives $\Gamma(N_1\to \psi\eta^{(*)})/H(M_1)$ (the ratio of
the partial decay width to the Hubble rate), just like 
$\tilde m_{\nu_1}/m_*$ does for the decays into $\nu h$.
The dark
sphalerons associated to the SU(2)$_h$ gauge interactions have the same effect as (an increase in) the Dirac mass term for 
$\Psi$ and therefore do not require additional consideration 
for the dark asymmetry. 

If there is no hierarchical structure to the couplings $y_\psi^{ij}$
and their phases are large, we can estimate 
$(y_\psi^\dagger y_\psi)_{kj}\sim
(\tilde y_\psi^\dagger \tilde y_\psi)_{kj}$
or its imaginary part by some
average value $\bar y_\psi^2$.  Further defining $\bar y_\nu^2 = 
(y_\nu^\dagger y_\nu)_{11}$ and assuming that the terms of
order $y_\nu^2$ in the numerator of (\ref{epsij}) can be estimated
as $\bar y_\nu^2$, we find that the CP asymmetry for $\psi$ is of order
\be
	\epsilon_{\psi 1} \sim {2\,\bar y_\psi^2\over 8\pi\,\sqrt{x}}\, 
	\left(1+\frac32 r\over 1+r\right)
\label{eps1est}
\ee
where we define $r = \bar y_\psi^2/ \bar y_\nu^2$, and assume that 
$\epsilon_\psi^{12}\sim - \epsilon_\psi^{11}$ in (\ref{Yieq}).  It is evident
that eq.\ (\ref{eps1est}) has only mild dependence upon $r$.
Combining with the efficiency factor $\kappa_\psi$ (where we approximate the
exponent $1.1$ by 1) leads to the estimate\footnote{\label{caveat}This is valid
for parameters such that $\kappa_\psi < 1$ hence $\tilde m \gtrsim 4 m_*$.
Using eq.\ (\ref{ynueq}) this implies $\bar y^2 \gtrsim 
10^{-7}(M_1/10^{10}{\rm\, GeV})$.  We will assume that this restriction
holds in the following.}
\be
	Y_\psi \cong 1.4\times 10^{-12}\left(M_1\over 10^{10}{\rm\,GeV}
	\right)\left(10\over x^{1/2}\right)
\label{Ypsieq}
\ee
ignoring $r$ dependence.

\subsection{Baryon asymmetry estimate}

We wish to explain the baryon asymmetry simultaneously with that of
dark matter.  Analogously to (\ref{Yieq}), it is given by
\be
	Y_B = {28\over 79}\cdot {45\over \pi^4}{\epsilon_{\nu,1} \kappa_\nu
	\over g_*}
\ee
where the prefactor $28/79$ is due to redistribution of the initial lepton
asymmetry into baryons via sphaleron interactions.   The CP asymmetry
$\epsilon_{\nu,1}$ is defined as $\epsilon_{\nu,1} =
\sum_i\epsilon_{\nu,1i}$ in the usual way for leptogenesis.
Similarly to our estimate in (\ref{eps1est}), we expect the 
well-known D-I bound \cite{Davidson:2002qv} to be modified by
a function of $r=\bar y_\psi^2/ \bar y_\nu^2$,
\bea
	|\epsilon_{\nu,1} | &\le& {3\over 16\pi}{M_1\over v^2}
	{\Delta m^2_{\rm atm}\over m_{\nu_3}}\, 
	\left(1+\frac23 r\over 1+r\right)\nonumber\\
	&\cong& 10^{-6}\left({M_1\,\over 10^{10}{\rm\, GeV}}\right)
	\equiv 10^{-6}\, M_{10}
\label{epsnu1}
\eea
where $\Delta m^2_{\rm atm} = m_{\nu_3}^2-m_{\nu_2}^2$, which we
assume to be $\cong m_{\nu_3}^2 \cong (0.05{\rm\, eV})^2$.
Again the dependence upon $r$ is mild, and 
we will ignore the effect of the DM Yukawa coupling on leptogenesis
in the visible sector.  To
estimate the efficiency factor 
$\kappa_\nu \cong 0.25\,(m_*/\tilde m_\nu)$, with 
$\tilde m_\nu = (y_\nu^\dagger y_\nu)_{11} v^2/M_1$, we use
the Casas-Ibarra parametrization of $y_\nu$,
\bea
	(y_\nu^\dagger y_\nu)_{11}  &\cong&
	U_{1i}\, m_{\nu_i}^{1/2}\, R^\dagger_{ik} {M_k\over v^2} R_{kj} 
	m_{\nu_j}^{1/2}\, U^\dagger_{j1}\nonumber\\
	&\cong& 10^{-6}\, M_{10}
\label{ynueq}
\eea
(the same result as eq.\ (\ref{epsnu1}))
where $U$ is the PMNS matrix and $R$ is an arbitrary SU(3)
transformation.   We assumed that $R^\dagger_{ik} {M_k} R_{kj}
\sim M_1$ since we take the heavy neutrino masses to be of the same
order, and $|U_{12}|^2 m_{\nu_2} + |U_{13}|^2 m_{\nu_3} = 0.003$ eV
(taking $m_{\nu_1}$ to be much less than the solar neutrino mass
splitting).  This gives
$\kappa_\nu \cong 1/12$ and 
\be
	Y_B \cong 1.4\times 10^{-10} \, M_{10}\, \epsilon_{\sss\rm DI}\,
\label{YBeq}
\ee
where we have introduced a parameter $\epsilon_{\sss\rm DI}$ to quantify how much
$\epsilon_{\nu,1}$ falls below the D-I bound, {\it i.e.,} 
$\epsilon_{\sss\rm DI}$
is $|\epsilon_{\nu,1}|$ over its maximum value.  Equating $Y_B$
to its measured value, we find $\epsilon_{\sss\rm DI}\, M_{10} = 0.7$.

\subsection{DM to baryon constraint}
We can combine the above results to get a constraint on the 
model parameters
from the known ratio of baryon and dark matter energy densities,
$\Omega_B/\Omega_{DM} = m_p Y_B / (m_{\bf H} Y_\psi) = 0.18$.  Here
$m_{\bf H} = m_1 + m_2$, the mass of the dark atom (neglecting its
binding energy).  Then we find
${m_{\bf H}/m_p} =  166\, \epsilon_{\sss\rm DI}\, (x/10)^{1/2}$.
We can eliminate $\epsilon_{\sss\rm DI}$ using eq.\ (\ref{YBeq}) and $M_1$
using (\ref{ynueq}) to obtain
\be
	{m_{\bf H}\over m_p} = 560\, \epsilon_{\sss\rm DI}\, 
	\left(x^{1/2}\over 10\right)
	= {360\over M_{10}} \left(x^{1/2}\over 10\right)
\label{mHeq}
\ee

Eq.\ (\ref{mHeq}) reveals part of our motivation for the choice $M_1
\sim 10^{10}\,$GeV: it gives dark atom masses in a range that is
interesting for direct detection and consistent with our prejudice for
the new physics scale to not be far below the weak scale.  It is
interesting that the same mass scale is also consistent with the
observed baryon asymmetry for generic choices of the neutrino CP
asymmetry, $\epsilon_{\sss DI} \lesssim 1$.  

Notably absent from our estimates is any explicit dependence upon the
Yukawa couplings $\bar y_\psi^2$ and $\bar y_\nu^2$.  This is because
of the cancellation between the CP asymmetry $\epsilon$ and the efficiency
factor $\kappa$, which only occurs for couplings such that 
$\kappa < 1$.  We verified this condition for $\kappa_\nu = 0.08$.
It is also satisfied by $\kappa_\psi$ so long as $\bar y_\psi^2 \gtrsim 
\bar y_\nu^2/12$.  We will make this technical assumption to keep
the analysis simple.  For smaller values of $\bar y_\psi^2$, there
would be a suppression of $Y_\psi$ and the need for correspondingly
larger values of $m_{\bf H}$.

As an example, we take $\epsilon_{\sss\rm DI} = 0.65$, $\bar y_\psi^2 =
\bar y_\nu^2 = 10^{-6}$, $m_{\bf H} \cong 83\,$GeV,
$M_1 = 10^{10}\,$GeV, $M_2 = \sqrt{x} M_1 = 2\times10^{11}\,$GeV.  
Larger or smaller values of $m_{\bf H}$ can be obtained by
adjusting $M_2/M_1^2$, using eq.\ (\ref{mHeq}).

\subsection{Decay of dark scalars}
\label{decays}

An interesting feature of our model is that the seesaw mechanism
produces the new dimension-5 operator (\ref{dim5op}) that allows the
dark scalars $\eta$ to decay \cite{Falkowski:2011xh}. When the SM Higgs takes its vacuum expectation value, this allows the $\eta$ to decay directly into $\nu\psi$. The decay rate is of order
\bea
\label{Gammaeq}
	\Gamma &\sim& \frac{\bar{y}_\psi^2 \bar{y}_\nu^2 m_\eta v^2}{8\pi M_1^2} \\
	&\sim&3\times10^{-3}\,\rm{s}^{-1}\cdot\left(\frac{m_\eta}{150\,\rm{GeV}}\right)
\eea
where for the numerical estimate of $\Gamma$ we used the
exemplary values specified at the end of the previous section
(ignoring the mass of $\psi$ in the phase space integral).

Such decays must occur sufficiently early so that the
decay products are fully thermalized before they can distort the CMB.
Ref.\ \cite{Slatyer:2012yq} shows that this occurs if the lifetime is
below $\sim 10^{12}$s.  Eq.\ (\ref{Gammaeq}) implies that 
our model easily satisfies this bound.

\begin{figure*}[t]
\centerline{
\includegraphics[width=\columnwidth]{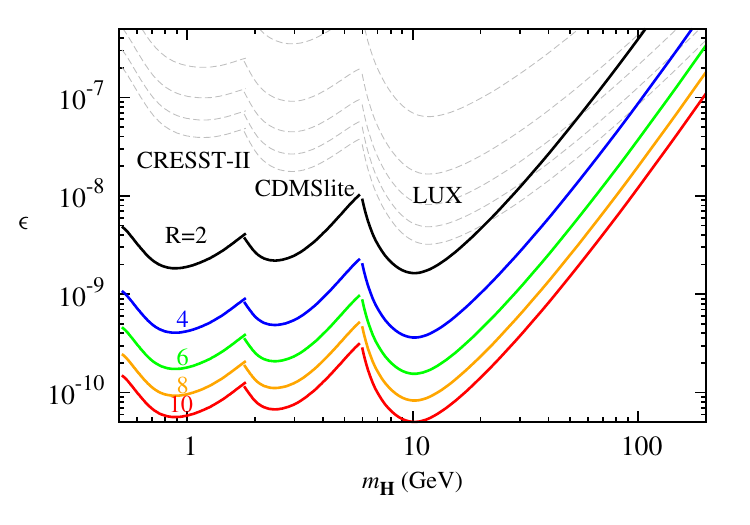}
\includegraphics[width=\columnwidth]{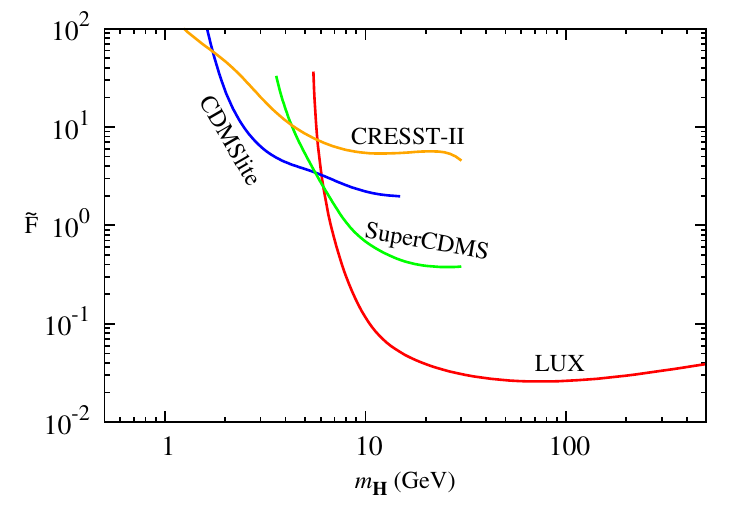}
}
\caption{Left: CRESST-II, CDMSlite and LUX
 limits on millicharge $\epsilon$ of dark atom
constituents, with constituent mass ratios
$m_2/m_1 = R = 2,\,4,\cdots,\,10$ as indicated,
for photon-mediated scattering of dark atoms on protons.
For clarity, only the most constraining limit is shown for any 
DM atom mass $m_{\bf H}$.
Gauge coupling is set to $\alpha_g = \alpha_{\rm ion}$, eq.\
(\ref{alpha_ion}) for solid curves, and fixed at $\alpha_g = 0.06$
for light dashed curves.  
Right: Corresponding limits on $\tilde F = y_1\,\theta |1-m_h^2/m_\phi^2|
F_\psi(0)$
from Higgs portal scattering, where $\theta$ is the $\phi$-Higgs
mixing angle.}
\label{fig:dd2}
\end{figure*}

\begin{figure*}[t]
\centerline{
\includegraphics[width=\columnwidth]{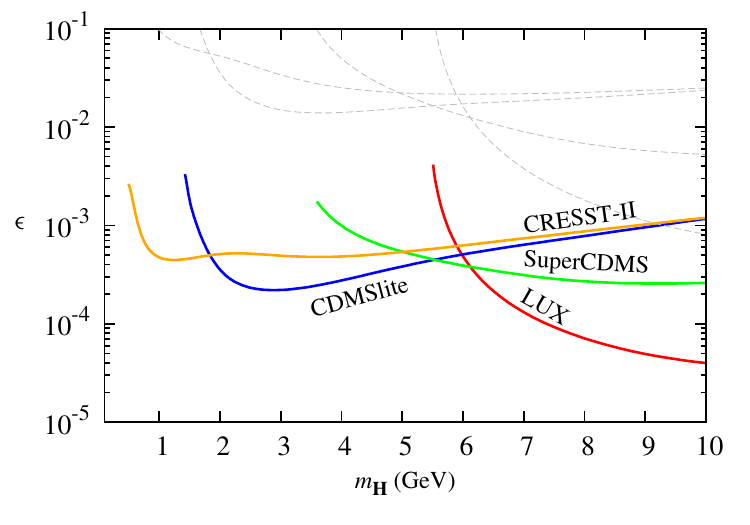}
\includegraphics[width=\columnwidth]{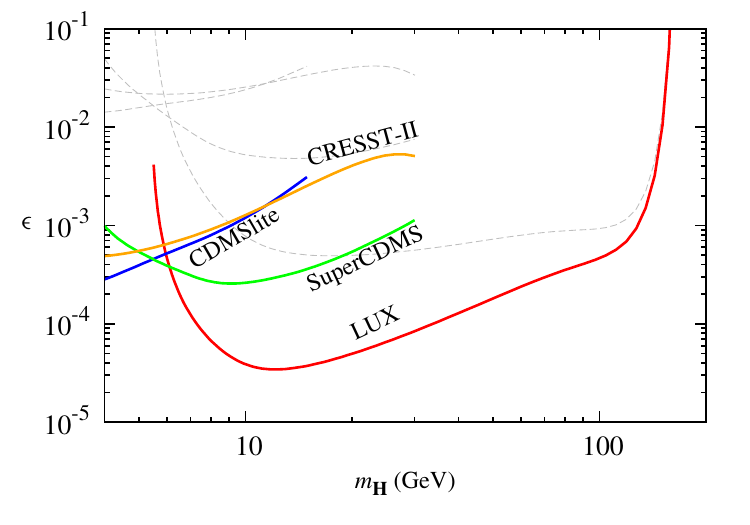}
}
\caption{Direct detection constraints on kinetic mixing
parameter $\epsilon$ versus dark atom mass $m_{\bf H}$ for case
of equal-mass constituents $m_1=m_2 = m_{\bf H}/2$, when interaction
is magnetic inelastic.  Hyperfine mass splitting is chosen as a
function of $m_{\bf H}$ as described in text.  Solid and dashed curves
refer to choice of $\alpha_g$ as in fig.\ \ref{fig:dd2}.
Left: low mass region;
right: larger mass region. }
\label{fig:dd1}
\end{figure*}

\section{Direct Detection}
\label{ddsect}

There are two portals through which our dark atoms to interact with nuclei.
The kinetic mixing allows for photon exchange, which has been 
discussed in refs.\ \cite{Cline:2012is,Cline:2012ei}.  The ensuing constraints on the electric
millicharge $\epsilon$ are weakened for atoms compared to ions because
of the screening of electric charge.  In the special case where
$m_{1} = m_{2}$ this screening is perfect, and the
interaction becomes magnetic dipole, further weakening the limits
\cite{Cline:2013zca}.
Here we extend
results of ref.\ \cite{Cline:2013zca} 
for the $m_{1} = m_{2}$ case to higher DM masses.

In addition, there is the Higgs portal induced by mixing of $h$ and
$\phi_3$ through the operator $\frac12\lambda_{h\phi} |H|^2|\phi|^2$.
$\phi_3$  interacts with the dark atom constituents through the
operator $\bar\Psi(y_1 + iy_2\gamma_5)(\vec\phi\cdot\vec\tau)\Psi$
that splits the $\Psi_{1,2}$ masses.  

\subsection{Kinetic mixing portal}

\subsubsection{Unequal-mass constituents}
As discussed in ref.\ \cite{Cline:2012is}, the mutual 
screening of the electric
charges of the $\Psi_1$ and $\Psi_2$ constituents results in a 
scattering matrix element where the $1/q^2$ of the photon propagator
is canceled by $q^2$ in the form factor for the charge density.
The cross section for scattering of dark atoms on a proton is
\bea
	\sigma_p &=&  
	  4\pi {\alpha^2\,\epsilon^2\, \mu_n^2\over 
	\alpha_g^4}\left({1\over m_1^{2}} -{1\over
	 m_2^{2}}\right)^2  \nonumber\\
	&=& 	  4\pi {\alpha^2\,\epsilon^2\, \mu_n^2\over 
	\alpha_g^4\, m_{\bf H}^4} \,f_0(R)
\label{sigmap}
\eea
where $\mu_n$ is the reduced mass of the dark atom and nucleon 
system, and $f_0(R) = (1+1/R)^4 (R^2-1)^2$.
Here we have
generalized the result of ref.\ \cite{Cline:2012is} where the
approximation of large $R$ was made.
The expression (\ref{sigmap}) is valid if $R$ is not too close to 1.
The question of ``how close?'' is discussed below.

For $R\neq 1$, the resulting upper limits on $\epsilon$ is 
illustrated in fig.\
\ref{fig:dd2}(a) showing the most constraining limit from 
the LUX \cite{LUX},
CRESST-II \cite{Angloher:2015ewa} or CDMSlite \cite{Agnese:2015nto}
experiments, at any given dark atom mass $m_{\bf H}$.  In section \ref{other} we will see that the requirement of
sufficiently small ionization fraction in the dark sector leads to the
constraint
\be
	\alpha_g \ge \alpha_{\rm ion} \equiv 5\times 10^{-3}
	\left(m_{\bf H}\over {\rm GeV}\right)^{1/2} f_2^{-1/4}(R)
\label{alpha_ion}
\ee
where 
\be
	f_2(R) = R + 2 + R^{-1}
\label{f2eq}
\ee
The solid curves are derived for 
the parameter choice which saturates this bound, $\alpha_g = \alpha_{\rm
ion}$, and $R$ ranging from 2 to 10,
while the dashed ones assume a fixed value of $\alpha_g= 0.06$.
This value satisfies the constraint $\alpha_g>\alpha_{\rm ion}$
over the entire range of $R$ and $m_{\bf H}$ shown on the plots.
(The unusual sensitivity of the solid curves to light DM masses is
due to the decrease of $\alpha_g=\alpha_{\rm ion}$ with $m_{\bf H}$, and
consequent increase in the dark Bohr radius, leading to larger
cross sections.)
The nominal constraints from the experiments are weakened by 
factors of $(A/Z)^2 = 5.9,\, 5.2$ and 4 respectively to account for 
the coupling to protons only.  For CRESST this corresponds
 to collisions with the oxygen atoms that dominate
the senstivity to low-mass dark matter. 
The strongest constraints occur for $m_{\bf H}\cong 1-10\,$GeV,
in the range $\epsilon \lesssim 10^{-10}-10^{-8}$.
For conventional abelian kinetic mixing, such small
values of $\epsilon$ could be difficult to achieve since the loop
diagram that generates it is not surpressed by any large mass scales,
since in this case the kinetic mixing operator is marginal.  However 
for nonabelian kinetic mixing, $\epsilon$ is suppressed by the 
mass $m_\chi$ of the heavy particle in the loop, as well as its
Yukawa coupling $y_\chi$.  For example if the couplings described
below eq.\ (\ref{ngkm}) are $y_\chi=0.1$, $g_1=g$,
$R=10$, $\alpha_g=\alpha_{\rm ion}$ and $\sigma = 30$ GeV, we require 
$m_\chi\gtrsim 3\times 10^{11}$ GeV to satisfy
the LUX bound on 10 GeV dark atoms.  

\subsubsection{Equal-mass constituents}
\label{req1}

For $R\cong 1$, there is perfect screening of charge because of the
complete overlap of the wave functions of the two constituents, and
the magnetic dipole interaction that we have neglected in
(\ref{sigmap}) becomes important.  This case was considered in detail
in ref.\ \cite{Cline:2012is}.  The magnetic scattering is inelastic 
because of the hyperfine transition of the dark atom, requiring
energy 	$\delta E = \frac16\alpha_g^4\, m_{\bf H}$, hence a minimum DM velocity
of $v_{\rm min} = q/(2\mu_N) + \delta E/q > \sqrt{2\,\delta E/\mu_N}$ 
for momentum transfer $q$ and dark atom-nucleus reduced mass $\mu_N$.
There is a $q$- and $v$-dependent form factor $F = (q_0/q)^2(v^2-v_{\rm
min}^2)/v_0^2$ that is of order unity for typical values
$v\sim v_0$ and $q\sim q_0$, as along as $v_0\gtrsim v_{\rm min}$.
The cross section on protons is of order
\be
	\sigma_{p,0} \equiv {64\pi\epsilon^2\alpha^2 \mu_n^2 v_0^2\over
	m_{\bf H}^2 q_0^2}
\label{sigmap0}
\ee
in that case, where $\mu_n$ is the proton-atom reduced mass.

More quantitatively, the actual cross section for a given scattering
event is $\sigma_p = \sigma_{p,0} F(q,v)$ and the
detection rate is proportional to 
\begin{align}
R&\propto Z^2\int_{E_{\rm{min}}}^{E_{\rm{max}}} 
dE_R\int_{v_{\rm{min}}}^{v_{\rm{esc}}}\frac{d^3\vec{v}}{v}
f(\vec{v})\,\sigma_{p}\\
&\propto Z^2\sigma_{p,0} I_{F},
\end{align}
with
\begin{align}
E_{\rm min} &= \frac12  m_{\bf H} v_{\rm min}^2\nonumber\\
E_{\rm{max}}&=p_{\rm{max}}^2/(2m_N)\nonumber\\
p_{\rm{max}}&=
\sqrt{\mu_n^2(v_{\rm{esc}}+v_0)^2-2\,\delta E\mu_N}+
\mu_N(v_{\rm{esc}}+v_0)\nonumber\\
f(\vec{v})&\propto e^{-(\vec{v}+\vec{v}_e)^2/v_0^2}-e^{-v_{\rm{esc}}^2/v_0^2},\\
I_F&\equiv\int_{E_{\rm{min}}}^{E_{\rm{max}}} 
dE_R\int_{v_{\rm{min}}}^{v_{\rm{esc}}}\frac{d^3\vec{v}}{v}
f(\vec{v})\,F(q,v).
\end{align}
Here $\vec{v}_e$ is the Earth's speed relative to the DM halo,
$v_0\approx220\,\rm{km}\,\rm{s}^{-1}$ is the mean DM velocity,
$v_{\rm{esc}}\approx450\,\rm{km}\,\rm{s}^{-1}$ is the approximate
escape velocity of the DM halo (we see no significant variation 
in the results for values in the range
$400-500\,\rm{km}\,\rm{s}^{-1}$), and $E_{\rm{esc}}$ is the maximum
recoil energy from a DM particle with the escape velocity.

We compare the rate for our model to that of
generic DM scattering with a constant cross section 
$\sigma_{n}$, for which the corresponding expressions are
\begin{align}
R&\propto A^2\sigma_{n}\,I_0,\\
I_0&\equiv\int_{E_{\rm{min}}}^{E^{(0)}_{\rm{max}}} dE_R\int_{v_{\rm{min}}}^{v_{\rm{esc}}}\frac{d^3\vec{v}}{v}f(\vec{v}).
\end{align}
where $E^{(0)}_{\rm{max}} = E_{\rm{max}}$ evaluated at ${\delta E=0}$.
Therefore the magnetic inelastic cross section (\ref{sigmap0}) is 
bounded from above as 
\begin{align}
\sigma_{p,0} < \frac{A^2\,I_0}{Z^2\,I_F}\,\sigma_{n,\rm{lim}}
\label{sigmap0lim}
\end{align}
where $\sigma_{n,\rm{lim}}$ is the experimental upper limit on
the cross section for a generic DM model.   Notice that the arbitrary
quantity $(v_0/q_0)^2$ appears in the same way on both sides of 
(\ref{sigmap0lim}) and hence can be divided out.

Although the gauge coupling $\alpha_g$ does not appear  in
(\ref{sigmap0}), the mass splitting $\delta E = \alpha_g^4 m_{\bf
H}/6$ depends upon it. For definiteness, we have chosen the value
$\alpha_g = \alpha_{\rm ion}$ in eq.\ (\ref{alpha_ion})
from the requirement of sufficiently small dark ionization fraction.
This fixes $\delta E$ as a function of $m_{\bf H}$.  

We plot the ensuing  limits on $\epsilon$ in figure \ref{fig:dd1},
using results from the  LUX \cite{LUX}, SuperCDMS 
\cite{Agnese:2014aze}, CRESST-II \cite{Angloher:2015ewa},
and CDMSlite \cite{Agnese:2015nto} experiments.  The mass
dependence of $\alpha_g = \alpha_{\rm ion}$ changes the shape of
the exclusion curves relative to those on the cross section itself,
since the mass splitting $\delta E$ rises rapidly with $m_{\bf H}$,
nullifying the signal for $m_{\bf H}\gtrsim 100\,$GeV. 

\subsubsection{Transition from $R=1$ to $R >  1$}

\begin{figure}[t]
\centerline{
\includegraphics[width=\columnwidth]{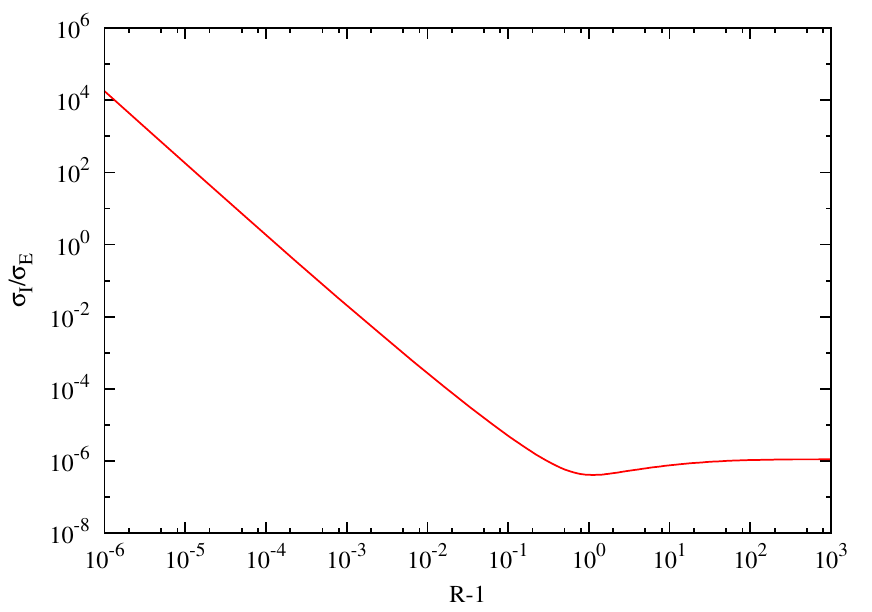}}
\caption{Ratio of the magnetic inelastic and elastic cross section for
scattering of dark atoms on protons as a function of the constituent
mass ratio $R$ (its deviation from unity), for dark atom mass
$m_{\bf H} = 10$ GeV and mass splitting described in section
\ref{req1}.}
\label{fig:dd3}
\end{figure}

We have noted that inelastic magnetic transitions dominate for
equal constituent masses, $R=1$, while elastic charge-charge
interactions dominate when $R > 1$.  One may wonder how sharp the
transition is between the two regimes; how small must $R-1$ be for
inelastic transitions to dominate?  We have calculated the ratio of
the two cross sections as a function of $R$ for a particular value of 
$m_{\bf H} = 10$ GeV as an example, taking the mass splitting $\delta
E$ as described above.  The result is graphed in fig.\ \ref{fig:dd3},
which shows that only for $R < 1.0001$ do the inelastic transitions
dominate.  Hence the most natural situation corresponding to this case
is that where $R=1$ exactly.  
There are two limits in our model that give $R=1$: either $y_1=0$, 
or $m_\psi = 0$.  The latter is a point of enhanced SU(2)
flavor symmetry for the two chiral (doublet) fermions.

\subsection{Higgs portal}
The interaction of dark atoms with the Higgs through $\phi_3$-$H$
mixing also undergoes screening
because of the coupling $\tau_3$ which has opposite sign for
$\Psi_1$ and $\Psi_2$.
At low
energies, the dark atoms can be described by a Dirac field $\bf H$
whose coupling to the virtual $\phi_3$ or $h$ carrying momentum $q$ is given
by the amplitude
\be
	y_1\,\bar u_{\sss\bf H} u_{\sss\bf H}\ F_\psi(q)
\label{Frule}
\ee
We have neglected the $y_2$ contribution that is suppressed by the
dark matter velocity.
By matching onto the scattering amplitudes in the
high-energy theory, we infer that
\be
	F_\psi(q) =  {1\over m_{\bf H}}\left[
	{m_2\over\left(1 + 
	\sfrac14q^2 a_2^2\right)^{2}} - {m_1\over \left(1 + 
	\sfrac14q^2 a_1^2\right)^{2}}\right]
\label{formfact}
\ee
with  $a_i = (\alpha_g m_i)^{-1}$.  Thus the coupling vanishes
in the limit $R=1$ ($m_1 = m_2$). 
If $\theta$ is the $h$-$\phi_3$ mixing angle, then the amplitude for
scattering of dark atoms on nucleons is
\bea
\label{ddamp}
	{\cal M} &=& y_1\,\bar u_{\sss\bf H}(p_3) u_{\sss\bf H}(p_1)
	\cdot \bar u_n(p_4) u_n(p_2)\!\left({y_n m_n\over v}\right)
	\\
	 &\times&
	c_\theta s_\theta\left({1\over m_h^{2}} 
	-{1\over m_\phi^{2}}\right)  F_\psi(q)
	\nonumber
\eea
where $(y_n m_n/v)$ with $y_n \cong 0.3$ \cite{scalarproton}
is the coupling of the Higgs to nucleons.

If $\alpha_g$ is not too small, we can take the $q=0$ limit of the form
factor.  In this case the cross section
for dark atom-nucleon scattering is 
\be
	\sigma_n \cong {1\over \pi v^2} 
	\left[y_1\,y_n m_n  \mu_{n\sss\bf
	H}\, \theta \,F_{\psi}(0)\right]^2 (m_\phi^{-2} - m_h^{-2})^2 
\ee
in terms of the $\bf H$-nucleon reduced mass, and taking $\theta\ll 1$.  The LUX upper limit on 
the dimensionless combination $\tilde F = y_1\,\theta |1-m_h^2/m_\phi^2|
F_\psi(0)$ is plotted in 
fig.\ \ref{fig:dd2}.  The Yukawa coupling $y_1$  is related to
the mass splitting in the dark sector since $m_2^2 - m_1^2 = 
y_1^2\sigma^2$.  Moreover it is straightforward to show that
$F_\psi(0) = (m_2^2 - m_1^2)/m_{\bf H}^2$.  If $m_\phi < m_h$
then $\tilde F \cong y_1^3\theta (m_h^2 \sigma^2)/(m_\phi^2 m_{\bf
H}^2)$.  We expect $m_\phi \sim \sigma$, similarly to $m_h \sim v$
in the visible sector, and $\theta\lesssim 0.01$ to satisfy LEP
constraints \cite{Schael:2006cr} on mixing of a light scalar with the Higgs.
The largest dark atom mass range for saturating the LUX bound shown in fig.\ 
\ref{fig:dd2} with $|y_1|\lesssim 1$  is $m_{\bf H} \lesssim 70$ GeV. 

\section{Stable Vector Bosons}
\label{h2sect}

Up to now we have assumed that the $\Psi_1$-$\Psi_2$ mass splitting 
is sufficiently small to prohibit the decay $\Psi_2^-\to
B^{--}\Psi_1^+$, corresponding to the condition
\be
	|y_1| < {m_1+m_2\over 4 m_\psi}\,g
\ee
However this need not be the case, and the model
is also compatible with a universe where charge neutrality in the dark
sector is acheived by having two $\Psi_1^+$ particles for every 
$B^{--}$.  This leads to a very different kind of dark atom that 
is reminiscent of the $H_2$ molecule, except that the two ``protons''
are bound together by a single charge $-2$ ``electron''.  We will 
refer to these variant dark atoms as ${\bf H_2}$.  In the absence of
fine-tuning, the stable vector
boson is typically lighter than $\Psi_1$, prompting us to define the
ratio 
\be
	R_2 = {m_1\over m_B} \ge 1
\ee
in analogy to $R = m_2/m_1$ for $\bf H$ atoms.\footnote{To get the
opposite situation where $m_B > m_1$, we need $(g\sigma)^2 > 
(m_\psi - y_1\sigma)^2 + (y_2\sigma)^2$.  This requires not only
$y_2$ to be small, but also an accidental cancellation between
$m_\psi$ and $y_1\sigma$.}

\begin{figure*}[t]
\centerline{
\includegraphics[width=\columnwidth]{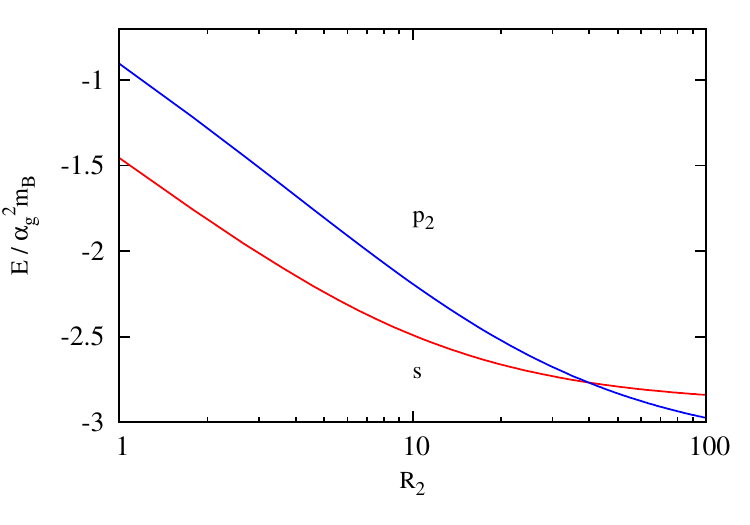}
\includegraphics[width=\columnwidth]{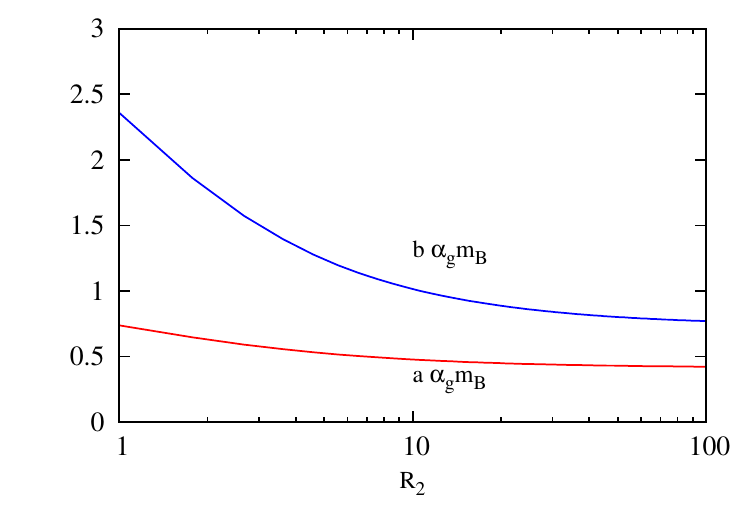}}
\caption{Left: energy obtained from variational method as a 
function of $R_2$ for the trial wavefunctions for 
$\bf{H_2}$ bound states $\psi_{{\bf H_2},s}$ and 
$\psi_{{\bf H_2},p2}$. 
Right: Corresponding values of the parameters $a,b$ (shown in 
dimensionless combinations with $\alpha_g m_B$) that determine
the spatial distributions of the wave functions, for the $s$-wave.}
\label{fig:contour}
\end{figure*}

\subsection{Bound States}

To verify the existence of the 3-body $\bf H_2$ bound states, we
make some ans\"atze for its wave function and use the variational method
to prove that the energy is minimized at a negative value.  
We consider trial wave functions where the positions of the three
particles are given by 
\be
	\vec x_\psi = \pm \vec\Delta/2, \quad \vec x_B = \vec r
\ee
{\it i.e.,} we work in the center-of-mass frame of the two $\Psi_1$
particles, with $\vec\Delta$ being their relative separation.
In analogy to the $H_2$ molecule, it could be expected that the
wavefunction for $\Delta$ is approximately that of a 3D harmonic
oscillator, $e^{-\Delta^2/b^2}$ for some scale $b$.  For simplicity
we take the wave function for $r$ to be hydrogen-like, $e^{-r/a}$
for some other scale $a$.  We consider three possible states,
an $s$-wave and two $p$-waves,
\begin{align}
\label{swave}
 \psi_{{\bf H_2},s}(\vec{r},\vec{\Delta})& = N_s \,
e^{-\Delta^2/b^2-{r}/a}\\
 \psi_{{\bf H_2},p1}(\vec{r},\vec{\Delta})& =  N_{p1} \,
r_z\,e^{-\Delta^2/b^2-{r}/a}\nonumber\\
 \psi_{{\bf H_2},p2}(\vec{r},\vec{\Delta})& =  N_{p2} \,
\Delta_z\,e^{-\Delta^2/b^2-{r}/a}\nonumber
\end{align}
where $r_z$ $(\Delta_z)$ is the $z$-component of $\vec r$
($\vec\Delta$).  

It is convenient to work in the analog of atomic units by rescaling
to dimensionless coordinates 
$r = r'/(\alpha_g m_B)$, $\Delta = \Delta'/(\alpha_g m_B)$.  Then
the Hamiltonian can be written as $H = (\alpha_g^2 m_B)H'$, where
the dimensionless $H'$ is 
\begin{align}
H' = -{1\over R_2}\nabla^2_{\Delta'} - \frac12\nabla^2_{r'} + {1\over \Delta'}
-\sum_\pm \frac{2}{|\vec{r}\,'\pm\vec{\Delta}'/2|}.
\end{align} 

By minimizing the expectation values $E = \langle \psi_{{\bf H_2}}|H|
\psi_{{\bf H_2}}\rangle$ with respect to $a,b$ and varying over a range of $R_2$
values, we find that bound states (having $E<0$) exist for all three
trial wave functions, but $\Psi_{{\bf H_2},p1}$ is always more weakly
bound than the other two.  Moreover the $s$-wave has lower energy
than $p2$ only for $R_2\lesssim 40$; for $R_2>40$ the $p2$ state is lower,
as shown in fig.\ \ref{fig:contour}.  Taking as an example the
values $m_1=60\,\rm{GeV}$, $R_2=10$, $\alpha_g=3\times10^{-2}$, the
three-constituent atoms have binding energies of approximately
$E\approx-15\,\rm{MeV}$.

\subsection{Direct Detection}

Dark $\bf H_2$ atoms interact similarly with nucleons relative to 
our treatment for $\bf H$ atoms in section~\ref{ddsect}, but there
are some qualitative differences, due to the more complicated wave
function.  In particular, there is no longer any special case
like $R=1$ for $\bf H$ atoms in which the electric millicharge clouds of the
constituents give exactly canceling contributions to the total
charge density.  This can be seen by computing the form factor, which
is the Fourier transform of the charge density
\bea
	\rho(x) &=& \int d^{\,3}\Delta\, d^{\,3} r\,
	|\Psi(\vec r,\vec\Delta)|^2\nonumber\\
	&\times&\left(
	\sum_\pm \delta(\vec x \pm \vec\Delta/2)  
	 - 2\delta(\vec x - \vec r)
	\right)
\eea
Using $\psi_{\bf H_2,s}$ from eq.\ (\ref{swave}), 
the form factor is
\begin{align}
F(q)&=2\left(-e^{-b^2q^2/32}+{1\over(1+a^2q^2/4)^{2}}\right)\nonumber\\
&\cong {q^2}\left(\frac{b^2}{16}-a^2\right),
\end{align}
where the approximation is for low momentum transfer $q$.

\begin{figure}[t]
\centerline{
\includegraphics[width=\columnwidth]{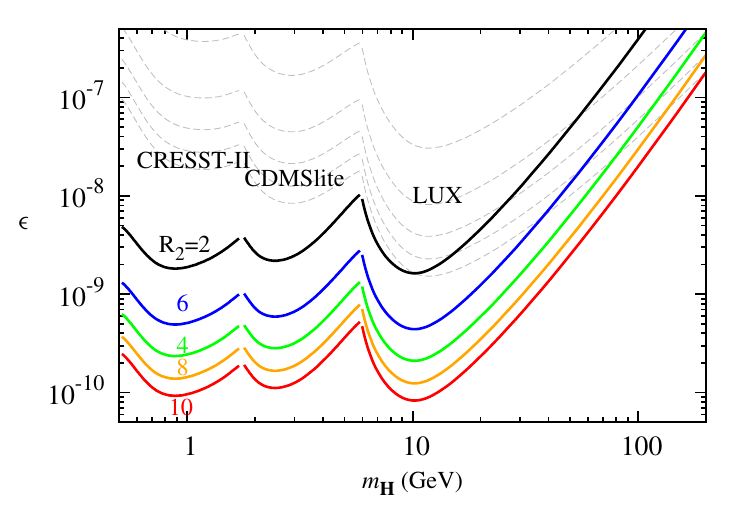}}
\caption{Direct detection constraints on kinetic mixing parameter 
as in fig.\ \ref{fig:dd2}, but for $\bf H_2$ atoms with
 $m_1/m_B \equiv R_2=2,4,\cdots,10$.}
\label{fig:LUXHe}
\end{figure}

In computing the cross section for scattering on protons, 
the factor of $q^2$ in the form factor cancels the $1/q^2$
of the propagator like before, giving
\begin{align}
\sigma_p=16\pi\alpha^2\epsilon^2\mu_n^2\left(a^2-\frac{b^2}{16}\right)^2.
\end{align}
at low momentum transfer. (The normalization can be deduced by
considering the limits $a=0,b\to\infty$ or {\it vice versa} where the
usual Feynman rules for the amplitude with no form factor apply.) The
direct detection limits from LUX \cite{LUX},  CRESST-II
\cite{Angloher:2015ewa} and CDMSlite \cite{Agnese:2015nto} through the
kinetic mixing portal are shown for various values of  $R_2$ in
figure~\ref{fig:LUXHe}, assuming  $\alpha_g$ saturates the constraint
(\ref{eq:h2ion}) from ionization of ${\bf H_2}$ atoms that we will
derive in the next section. (We also show the constraints for the
fixed value of $\alpha_g = 0.01$ as dashed curves.)  Unlike
with the dark atoms, the form factor never vanishes for any value of
$R_2$ (since $b$ is always $< 4a$).

For the Higgs portal, we follow the procedure in section~\ref{ddsect}. The amplitude and cross section are
\begin{align}
\mathcal{M}&=\bar{u}_{\bf H_2}(p_3)u_{\bf H_2}(p_1)
\cdot\bar{u}_n(p_4)u_n(p_2)\left(\frac{y_nm_n}{v}\right)\nonumber\\
&\times\left(\frac{c_\theta s_\theta}{m_h^2}-
\frac{c_\theta s_\theta}{m_\phi^2}\right)
\left(y_1\,F_\psi(q)+g\,F_B(q)\right)\\
\sigma_n&\cong\frac{1}{\pi v^2}\left[\left(y_1F_\psi(0)+g\,F_B(0)\right)
y_nm_n\mu_n\theta\,\right]^2\nonumber\\
&\times(m_\phi^{-2}-m_h^{-2})^2.
\end{align}
We have again made the approximation $\theta\ll1$ and assumed a 
small momentum transfer. $\mu_n$ is the $\bf{H_2}$-nucleon reduced 
mass, and $y_n\cong 0.3$ is the Higgs coupling to nucleons (modulo
$m_n/v$). The form factors are given by
\begin{align}
F_\psi(q)&=\frac{2m_\Psi}{m_{\bf H_2}}e^{-b^2q^2/32}\nonumber\\
F_B(q)&=\frac{m_B}{2m_{\bf H_2}}\frac{1}{(1+\sfrac14a^2q^2)^2}.
\end{align}
Redefining $\tilde{F}=(y_1F_\psi(0)+g\,F_B(0))\theta|1-m_h^2/m_\phi^2|$,
the constraint on $\tilde{F}$ from the LUX, CRESST-II and CDMSlite
experiments takes the same form as was previously shown 
shown in figure~\ref{fig:dd2} (right), where $m_{\bf H}$ is
reinterpreted as $m_{\bf H_2}$.   

\subsection{Neutron Star Constraints}

Tight constraints exist on the  cross section for asymmetric bosonic
dark matter scattering on nucleons from the existence of long-lived
neutron stars~\cite{Goldman:1989nd,McDermott:2011jp}. If the rate of
dark matter accretion is large enough, it can collapse to form a black
hole that would consume the progenitor, on time scales shorter than
the ages of neutron stars observed in globular clusters. In our model
it is important that we have only  one kind of stable bosonic dark
matter consituent carrying dark U(1)$_h$ charge. In the case of
$\rm{\bf{H}}$ atoms with only fermionic constituents, the  would-be
scalar constituents decayed early in the cosmological history, leaving
no asymmetric scalars. For ${\bf H_2}$ atoms, on the other hand,  the
vector bosons are mostly bound inside of atoms that resist collapse
because of the degeneracy pressure of their fermionic constituents.
The ionized fraction also resists collapse because of dark Coulomb
repulsion.  In contrast, in a model containing two species of bosons
carrying different U(1)$_h$  charges, nothing would prevent the
collapse of the combined bosonic fluid.

In more detail, we first note that the dark atoms remain bound once they start
to accumulate in the neutron star. From figure~\ref{fig:contour}, the binding energy is given by
\begin{align}
E_b \approx 2\,\alpha_g^2\,m_B = {2\,\alpha_g^2\,m_{\bf H_2}\over
	1+2R_2}
\end{align}
Using the dark ionization
constraint (\ref{eq:h2ion}), we find that $E_b > 130\,$eV even for the
extreme parameter choices $m_{\bf H_2} = 1$ GeV, $R_2=100$, which
is higher than
the temperature of the star, of order $100\,\rm{eV}$ 
\cite{Zurek:2013wia}.
Moreover fermions within a neutron star are supported by their 
degeneracy pressure, given by
\begin{align}
p=\frac{(3\pi^2)^{2/3}}{5\,m_{\psi}}n_{\psi}^{5/3},
\end{align}
where $n$ is the number density. 
A larger fermion mass decreases the pressure, and therefore the dark
atoms will tend to remain bound.

As for any ionized bosons that accumulate within the neutron star,
their repulsive self-interaction greatly weakens the bounds
on scattering with nucleons by preventing their collapse into a black hole.   Ref.\ \cite{Kouvaris:2011fi} 
finds that a repulsive scattering cross section exceeding 
$10^{-50}\,$cm$^2$ is sufficient to avoid neutron star constraints
for $m_{B} < 1$ TeV.  In our case the cross section corresponding
to dark Rutherford scattering is infrared divergent, but if we make it
finite by multiplying $d\sigma/d\Omega$ by $(1-\cos\theta)^2$ (thus
taking into account only scatterings with significant momentum
transfer), it is of order $\alpha_g^2/m_{B}^2 \gtrsim
10^{-34}\,$cm$^2$, where we used (\ref{eq:h2ion}) and $m_{B}\lesssim 100\,$GeV.  This satisfies the requirements of 
\cite{Kouvaris:2011fi} 
by many orders of magnitude.

\section{Other constraints}
\label{other}

Dark atoms, dark matter with millicharges, and models with  asymmetric
dark bosons are subject to further constraints from cosmological,
astrophysical and laboratory probes.  Here we discuss those coming from 
dark recombination, 
self-interactions of the dark matter and accumulation in neutron
stars, and searches for millicharged particles.

\subsection{Dark ions}
If the constituents of the hidden sector fail to combine into
atoms, they can scatter very strongly with each other through the
dark Coulomb interaction, contradicting the normally assumed
properties of collisionless cold dark matter.  From fitting to results of ref.\ 
\cite{Kaplan:2009de}, one finds that the ionization fraction can be
estimated as \cite{Cline:2012is,Cline:2014eaa,CyrRacine:2012fz}
\bea
\label{eq:ionization}
X_e &\cong& \left(1+10^{10}f_2(R)\,\xi^{-1}\,\alpha_g^4
	\frac{\rm{GeV}^2}{m_{\bf H}^2}\right)^{-1}
\eea
where $f_2(R) = R + 2 + 1/R$ (introduced in eq.\ (\ref{f2eq})), and $\xi$ is the ratio of dark sector to SM sector temperatures.

In \cite{Kaplan:2009de} it was argued that observations of the Bullet
Cluster rule out $X_e\gtrsim 0.1$, leading to the conservative lower limit 
$\alpha_g > \alpha_{\rm ion}$ 
(\ref{alpha_ion}) that we already incorporated in our analysis
of direct detection constraints. 
Ref.\ \cite{CyrRacine:2012fz} estimates that there is a factor of
10 uncertainty in (\ref{eq:ionization}).  We note that
this leads to only a factor of $1.8$ uncertainty in the expression 
for $\alpha_{\rm ion}$.

The ratio between temperatures can be found using the relation \cite{CyrRacine:2012fz}
\begin{equation}
\xi=\left(\frac{g_{*S,SM}^{0}\,g_{*S,D}^{\rm dec}}{g_{*S,SM}^{\rm dec}\,g_{*S,D}^0}\right)^{1/3},
\end{equation}
with $g_{*S,SM}$ and $g_{*S,D}$ denoting the number of degrees of freedom in 
the visible and dark sectors, and the superscripts
$0,{\rm dec}$ indicating the respective values today and at the time the
two sectors decouple kinetically. 
The temperature at which this decoupling occurs is therefore relevant.
We find that mixed Compton scattering with one dark and one SM photons
is the most important process for maintaining kinetic equilibrium.
It goes out of equilibrium when $H=n_\gamma\langle\sigma v\rangle$,
leading to the estimate
\begin{align}
1.66\,g_*\frac{T^2}{m_{\rm Pl}}&\sim g_*T^3\frac{8\pi}{3}\frac{\epsilon^2\alpha^2}{m_{\bf H}^2},
\end{align}
Thus mixed Compton scattering keeps the two sectors at the same 
temperature until
\begin{align}
T_{\rm dec}=
\,{3\times10^{-6}\,{\rm eV}\over \epsilon^2}\left(\frac{m_{\bf H}}{\rm{GeV}}\right)
\end{align}

The lowest value of $T_{\rm dec}$ is obtained by saturating the
direct detection limits on $\epsilon$ as a function of $m_{\bf H}$,
as shown in figs.\ \ref{fig:dd2}-\ref{fig:dd1}.
In the case of $R=1$ (equal mass dark atom constituents), 
this can be much
lower than the dark recombination temperature $T_{\rm rec}$, 
so that in fact
$T_{\rm dec}=T_{\rm rec}$, since Compton scattering is no longer
efficient on neutral atoms.
For $R>1$ on the other hand, the constraints on $\epsilon$ are 
sufficiently strong that the decoupling temperature
is limited to $T_{\rm dec}>300\,{\rm TeV}$. 

As long as $T_{\rm dec}\gg1$ TeV, all particles are
relativistic except for the heavy neutrinos. We therefore use the
values $g_{*S,SM}^{0}=3.94$~\cite{Cyr-Racine:2013fsa}, $g_{*S,SM}^{\rm
Dec}=106.75$, $g_{*S,D}^0=2$, and $g_{*S,D}^{\rm Dec}=18$. The
resulting temperature ratio is $\xi\approx 0.71$. At the other
extreme, decoupling occurs after electrons have frozen out. This
corresponds to $g_{*S,SM}^{\rm Dec}=7.25$, $g_{*S,DM}^{\rm Dec}=2$,
and $\xi\approx0.81$. Even at the two extremes, therefore, the
difference is minimal, and is further mitigated by the fact that $\xi$
is raised to the $1/4$ power in calculating $\alpha_{\rm ion}$. We
therefore adopt the value $\xi=0.71$ in eq. (\ref{alpha_ion}) so that
$\alpha_{\rm ion}$ remains a reasonable lower limit for $\alpha_g$.

There are certain cases that can lead to a lower temperature ratio,
with the smallest being that in which all dark content apart from the
dark photon has frozen out prior to the freeze-out of the top quark,
with decoupling occurring some time between these; in this case the
dark temperature could be as low as 0.3. These cases, however, are
unrepresentative and only apply to a narrow range of values of
$\epsilon$. Even in the extreme case of $\xi\approx0.3$, the estimate
on $\alpha_{\rm ion}$ would only differ by a factor of $\approx0.8$,
which is smaller than the error due to the uncertainty in the ionization
fraction.

\subsubsection{${\bf H_2}$ ionization}

For the case where $\Psi_2$ can decay to $\Psi_1$ and the vector boson
$B$, to make a rough estimate of the ionization fraction,  we assume
that recombination will typically happen in two steps: in the first,
unbound $\Psi_1$'s combine with the free $B$'s to make a $\Psi$-$B$
ion, while in the second these
ions bind with a second $\Psi_1$.  The first step is
similar to hydrogen atom recombination with the substitution
$\alpha_g\to 2\alpha_g$ due to $B$ having charge 2. 
In the second step, the potential at long
range is like that for hydrogen atom recombination. Equation
(\ref{eq:ionization}) then becomes
\bea
X_{e1}&\cong&\left(1+\xi^{-1}\,16\times10^{10}\alpha_g^4
\frac{\rm{GeV}}{m_1 m_B}\right)^{-1}\nonumber\\
X_{e2}&\cong&\left(1+\xi^{-1}\,10^{10}\alpha_g^4
\frac{\rm{GeV}}{m_{1}(m_B+m_1)}\right)^{-1}\nonumber\\
X_{e,{\rm tot}}&=& X_{e1}+X_{e2} \cong X_{e2}
\eea

The constraint on the ionization fraction ($X_{e,{\rm tot}}\lesssim
0.1$) from 
\cite{Kaplan:2009de} is therefore
\begin{align}
\label{eq:h2ion}
\alpha_g\gtrsim \xi^{1/4}\,4\times10^{-3}
\left(\frac{m_{\bf H_2}}{\rm{GeV}}\right)^{1/2}f_3^{-1/4}(R_2)
\end{align}
where $f_3(R_2) = (R_2 + 1/2)^2/(R_2+R_2^2)$.

\subsection{Self interactions}

Although standard cold dark matter is considered to be noninteracting
with itself, there has been interest in variant theories where dark
matter has an elastic self-scattering cross section of order 1b per
GeV of DM mass.  This has been motivated by persistent discrepancies
between predictions of $N$-body simulations and observed properties of
dark matter halos.   While simulations tend to predict cuspy density
profiles for galaxies, there is some observational evidence for cored
profiles, especially in dwarf spheroidals.  Simulations also tend to
predict too many high-mass satellite galaxies accompanying
Milky-Way like progenitors compared to observations. For a review of these problems and their
possible resolutions, see ref.\ \cite{Weinberg:2013aya}.  A number of studies have
been done indicating that the small-scale structure problems can be 
alleviated by invoking dark matter elastic scattering with
$\sigma/m\sim 1$b/GeV.  Dark atoms can naturally accommodate such
large cross sections since they can have a significant geometric size.

The elastic scattering of dark atoms on each other has been studied 
very quantitatively, thanks to the fact that the problem can be mapped
onto that of normal atom scattering with appropriate rescalings of
parameters \cite{Cline:2013pca}.  A useful rough estimate is that the
scattering cross section goes as $\sigma \cong 100\, a_0'^2\cong 
100\, \alpha_g^{-2}\,f_2^2(R)\, m_{\bf H}^{-2}$.  A cosmologically
interesting level of self-scattering requires $\sigma /m_{\bf H} \sim
1.1\,$b/GeV $\cong 2800\,$GeV$^{-3}$
\cite{Zavala:2012us} in order to address the structure formation
problems of cold dark matter.  This corresponds to a gauge coupling of
\be
\alpha_g = 0.2\,  f_2(R)(m_{\bf H}/{\rm GeV})^{-3/2}
\label{alpha_int}
\ee

The criterion (\ref{alpha_int}) can be 
compatible with the ionization constraint (\ref{alpha_ion})
if $m_{\bf H}$ is sufficiently small, 
\be
	m_{\bf H} \lesssim 14{\rm\ GeV}\left(f(R)\over 4\right)^{5/8}
\label{mhbound}
\ee
obtained from eliminating $\alpha_g$ from the two relations.
Very large values of $R$ would be unnatural in our model, since it
would require a fine-tuned cancellation between two contributions
to $m_1^2 = (m_\psi - y_1\sigma)^2 + (y_2\sigma)^2$, as well as 
a small value of $y_2$.  An accidental cancellation at the level of
$R=10$ would allow for $m_{\bf H}$ as large as 28 GeV.

\subsubsection{$\bf H_2$ self-interactions}

In the $\bf H_2$ phase of the theory, the size of the atom is
determined by the length scale $a$ that describes the vector boson
part of the wave function, rather than the characteristic distance
$b$ between the fermions, even though $b \sim 2a$.  This is because
the expectation values are $\langle r\rangle = 1.5\, a$, $\langle
\Delta/s\rangle = 0.4\, b$.   Therefore in parallel to the $\bf H$
atom case, we can estimate the elastic cross section for atom-atom
scattering as $\sigma \cong 100\, a^2
\cong
100\,\alpha_g^{-2} m_{\bf H_2}^{-2} R_2^2\,f_4^2(R_2)$, where
$f_4(R)=1+(2R_2)^{-1}$.

The gauge coupling corresponding to the desired scattering cross section of $\sigma/m_{\bf H_2}=1.1\,\rm{b}/\rm{GeV}$ is therefore
\begin{align}
\alpha_g=0.19\, R_2\,f_4(R_2)\left(\frac{m_{\bf H_2}}{\rm{GeV}}\right)^{-3/2}.
\end{align}
When combined with the constraint (\ref{eq:h2ion})
on the ionization fraction, the result is
\begin{align}
m_{\bf H_2}\lesssim 6.9\,{\rm GeV}\,R_2\,f_4\, f_3^{-1/4},
\end{align}
which is similar to the expression found for the $\bf H$ case.  The
primary difference here is that large values of $R_2$ can be obtained
without fine-tuning of model parameters, allowing for a larger natural range of masses
consistent with both the ionization fraction and self-interaction
constraints.  (Notice that $f_{3,4}\to 1$ as $R_2$ becomes large.)
Even with a moderate hierarchy $R_2=10$, we can reach masses as large
as $m_{\bf H_2}\sim 70\,$GeV.

\subsection{Laboratory millicharge searches}

\begin{figure}[t]
\centerline{
\includegraphics[width=\columnwidth]{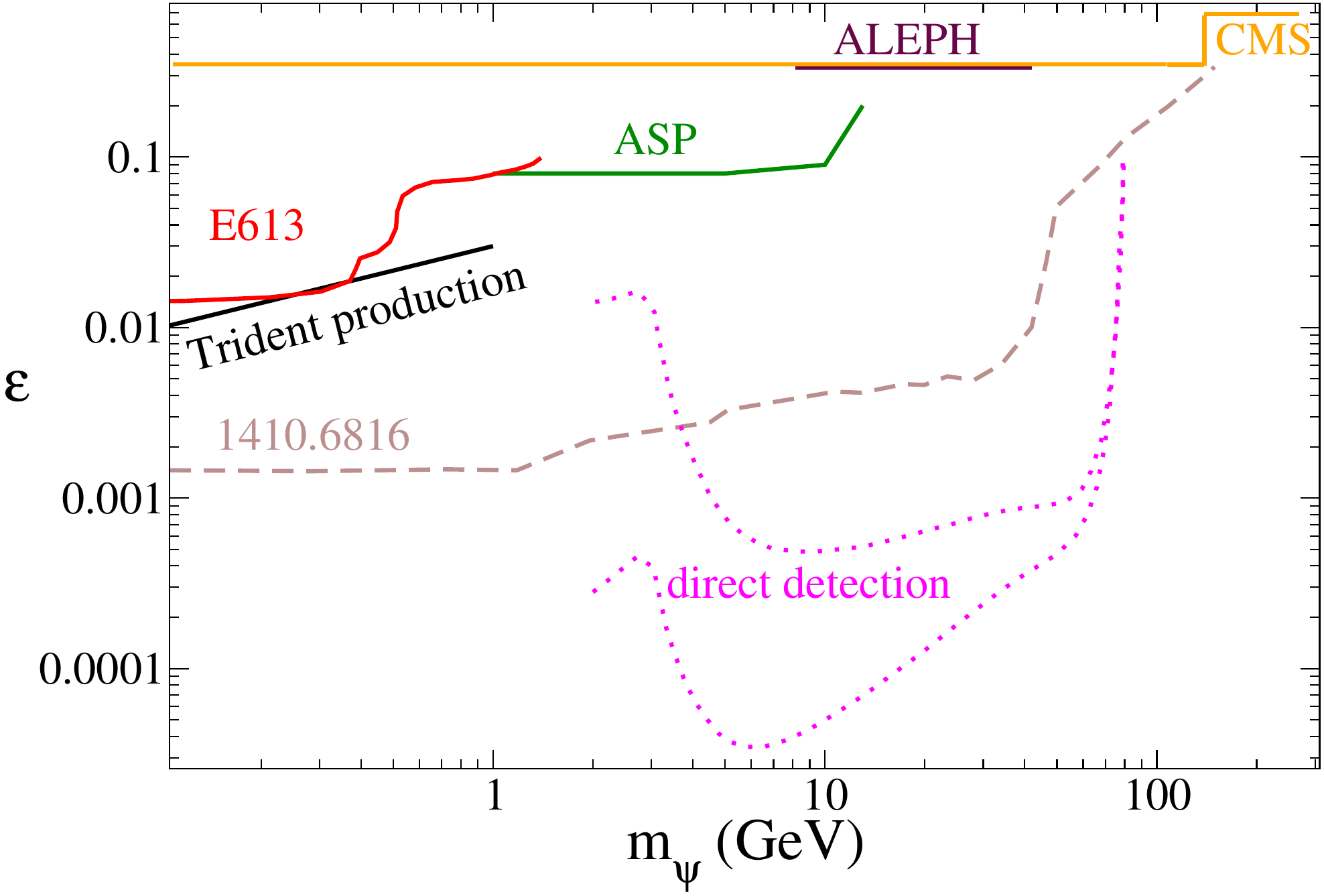}}
\caption{Solid curves: existing collider constraints on millicharge
versus mass; dashed curve: expected reach of experiment proposed
in ref.\ \cite{Haas:2014dda}.  Dotted curves: our direct detection
limits from fig.\ \ref{fig:dd1}, depending on choice of $\alpha_g =
0.06$ (upper curve) or $\alpha_g = \alpha_{\rm ion}$ (lower).}
\label{fig:collider}
\end{figure}

Pair production of $\bar\Psi_i\Psi_i$ is possible in accelerator
experiments from the coupling of the photon to the dark matter
millicharge.  The resulting constraints on $\epsilon$ are quite
weak in the mass range relevant for our model, $m_{\bf H} \sim 1-100$
GeV, as we show in fig.\ \ref{fig:collider}.  The existing constraints
are taken from tables in ref.\ \cite{PhysRevD.43.2314} for the
ASP and trident production limits, the E613 beam dump limit
\cite{PhysRevD.35.391}, ALEPH limits on the $Z$ decay width 
\cite{Buskulic:1992mr} and a  recent CMS search for particles
of charge $1/3$ or $2/3$ \cite{CMS:2012xi}.  We also show the reach of
a new proposed experiment for LHC (dashed curve) 
\cite{Haas:2014dda}.   These
constraints are considerably weaker than that coming from direct
detection, fig.\ (\ref{fig:dd1}), which is replotted as the dashed
curve on fig.\  \ref{fig:collider}.  Only at low ($m_{\bf H} \lesssim
4\,$GeV)
or high ($m_{\bf H} < 100\,$GeV) masses, outside the sensitivity of
direct detection, do they become dominant.

Possibly more significant constraints on millicharged particles arise
from searches for exotic isotopes, bound states of normal nuclei with
the charged DM constituents.  Very stringent limits on the
concentration of heavy isotopes of hydrogen or oxygen from sea water
have been derived; for example ref.\ \cite{Smith:1982qu}) obtains
an upper bound of $10^{-28}$ for the concentration of anomalously
heavy H. These experiments assume integer-charged ions, but a
recent experiment geared toward millicharged particles with $\epsilon >
10^{-5}$ set a limit of $10^{-14}$ on the abundance per nucleon.  
Naively such results would seem to rule out almost any values of
$\epsilon \gtrsim 10^{-3}$ such that the binding energy $E_b \cong
\frac12(\alpha\epsilon)^2 m_p$ (for anomalous H) exceeds $kT$ at room temperature, since
we expect some fraction of $\psi$ particles to remain ionized and thus
be able to contaminate normal matter.  

However to translate these limits on abundances into bounds on
$\epsilon$  requires many considerations, including the expected flux
of $\psi$ particles, their capture cross section on the elements in
question, the shielding of the earth and the galaxy from charged
particles by magnetic
fields, expulsion of charged particles by supernova winds, the process
of purification of the samples studied, and the question of whether
they apply to noninteger charged isotopes \cite{Cline:2012is}.  A
recent study of these issues was presented in ref.\
\cite{Kouvaris:2013gya}.  Here we take the view that there may be
room for evading the anomalous isotope searches, but this question should
be revisited if positive evidence for millicharges is found.

\section{Conclusion}
\label{conclusion}

In this work we have tried to strike a balance between simplicity and
realism in the construction of an atomic dark matter model.  Our
nonabelian construction is sufficiently rich to explain a unified origin of
the massless dark photon and charged (under the hidden U(1)$_h$
interaction) DM constituents $\Psi_i$ as a consequence of symmetry breaking
SU(2)$_h \to$ U(1)$_h$ by a scalar triplet VEV in the dark sector.  With the addition of 
a dark Higgs doublet, we have the necessary ingredients to explain the
$\Psi_i$ asymmetry through leptogenesis, simultaneously with the
baryon asymmetry.  Electric millicharges of $\Psi_i$, while not a
necessary ingredient, can arise naturally through heavy states
carrying both electric and U(1)$_h$ charge.  Higgs portal interactions
are also optional, but are allowed by a dimension-4 interaction of 
$\Psi_i$ with the dark Higgs triplet and its mixing with the SM Higgs.

The model is mainly testable by direct  
detection. For
sufficiently light or heavy constituents, the DM could also be discovered
in an experiment proposed for LHC to probe millicharged  particles. 
It can accommodate strong DM self-interactions as suggested by
problems of $\Lambda$CDM simulations to correctly predict the
small-scale structure of galaxies, if the dark atoms are not too
heavy.  Because of the requirement $\alpha_g \gtrsim \alpha_{\rm
ion}$, needed to make the ionization fraction in the dark sector
sufficiently small, the symmetric component of the dark matter is
highly suppressed due to annihilations into dark photons, making
any indirect signals too weak to be detected.

Our model has a number of features that distinguish it from  simplified
atomic dark matter models.  For example in the latter, the ratio $R$ of the masses
of the atomic constituents (which plays an important role) can be
arbitrarily large, whereas here it is naturally of order 1, and
requires fine-tuning to be much greater.

If the new Yukawa coupling $y_1$ exceeds the gauge coupling $g$, the
stable dark matter particles can be the lighter fermion $\Psi_1$ and the
doubly charged (under U(1)$_h$) vector boson $B^{--}$, leading to
novel three-body $B\Psi\Psi$ bound states, where the mass ratio of the
constituents $m_1/m_B$ could be large without tuning of parameters
(other than the usual hierarchy problem of light bosons).  The
properties of these unusual atoms for direct detection,
as well as for DM self-interactions, are qualitatively similar to those
of the more conventional two-constituent atoms.  This demonstrates
a loophole for strong neutron star constraints on asymmetric bosonic
dark matter, since the dark Coulomb repulsion prevents Bose
condensation in this model.

For future work, these models suggest a potential novel signal for 
direct detection, due to the possible simultaneous presence of  both
dark atoms and a subdominant component of ionized or symmetric
constituents.  This would allow for the detection of both types of
dark matter, typically having similar but distinct masses and
interaction cross sections.   Our analysis of leptogenesis as a common
origin of the visible and hidden asymmetries is approximate, and it
might also be interesting to undertake a more refined treatment for
future studies.

{\bf Acknowledgments}.  We thank Sacha Davidson, Kimmo Kainulainen,
Tim Linden, Zuowei Liu, Wei Xue and Wells Wulsin
for helpful correspondence or discussion.  We acknowledge support
of the Natural Sciences and Engineering Research Council of Canada.
We are grateful to NBIA
for its generous hospitality while we were completing this work.

\bibliography{boundstatebib}{}
\bibliographystyle{hep}

\end{document}